# 3D metamaterial with stiffness and Poisson's ratio approaching negative infinity

Zian Jia[1], and Lifeng Wang[1, *]

[1]*Department of Mechanical Engineering, State University of New York at Stony Brook, Stony Brook, New York 11794, USA*

We report a three-dimensional mechanical metamaterial that simultaneously possesses negative stiffness, negative bulk modulus, and negative Poisson's ratio. This metamaterial is a periodic arrangement of binder-shell elements. Under compression, the spherical shells dent inwards which cause the material to contract in the lateral directions. At larger deformations, snap through instability occurs and the material exhibits negative incremental stiffness. Interestingly, both incremental stiffness and incremental Poisson's ratio approach negative infinity (under displacement control) when snap-back is observed. We further showed that the multi-negative index metamaterial satisfies the strong ellipticity condition, therefore, a block of the metamaterial with many unit cells is stable under displacement constraint.

Young's modulus ($E$), shear modulus ($G$), bulk modulus ($K$), and Poisson's ratio ($\nu$) are fundamental material parameters that characterize a material's mechanical performance. Conventional materials have positive material parameters, and materials with negative parameters are a class of metamaterials [1-8]. A material with negative Poisson's ratio (auxetic materials) contracts laterally when compressed [4,9,10], while a material with negative stiffness deforms in the direction opposite to the applied force [3,11]. Negative indices are of great interest not only for their rich physics but also for their novel applications and exceptional mechanical performances. For instance, auxetic materials can serve as mechanical diodes (an auxetic material stuck in a tube is easy to push in but hard to pull out) and produce composites with unbounded shear to bulk modulus [12] and high crashworthiness [13,14]. Negative stiffness materials, on the other hand, can provide high energy absorption [15] and produce composites with unbounded damping and stiffness [3,16,17]. In recent decades, negative Poisson's ratio has been demonstrated in many materials [4,18-38], but 3D isotropic or quasi-isotropic materials are limited. Negative stiffness, by contrast, is acknowledged in material softening/failure [39], resonant systems [40,41], ferroelastic materials undergoing constrained phase transformation [42], and prestressed buckling [43]. Recently, bistability is shown in several metamaterials, which provides a promising way to rationally control negative stiffness. Yet, 3D metamaterials simultaneously possess negative stiffness, negative Poisson's ratio are largely unexplored.

Physically, unconstrained materials (with stress boundary condition) are stable if their strain energies are positive definite, which entails positive $E$, $G$, and $K$, and narrows the Poisson's ratio to $-1 < \nu < 0.5$ for isotropic materials [3,44] and $\nu < 0.5$ for cubic materials. Constrained materials (surface displacement boundary condition), in contrast, has a looser stability condition, which requires the material's elasticity tensor to be strongly elliptic, such that deformations are not localized [3,12,45]. The strong ellipticity condition entails $G > 0$ and $E(1-\nu)/\left[(1+\nu)(1-2\nu)\right] > 0$ [12]. In the case of $E < 0$, it entails $\nu < -1$ [12]. For cubic materials, strict verification of the elliptic condition requires checking the wave speed in all the wave propagation directions as

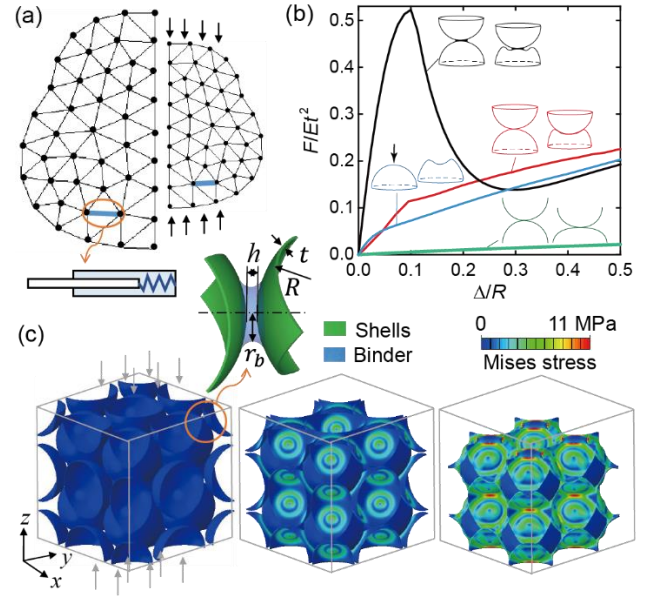

FIG. 1 (a) A "sleeved slider unit" network produces isotropic negative Poisson's ratio. The design concept is adopted from [46]. (b) Normalized force-displacement curves and deformation patterns of several structural elements (── one sphere, ── two spheres in contact, ── two cylinders in contact, and ── two spheres connected with a binder) under compression. (c) Compression of a metamaterial produced by periodically arranging the binder-shell elements in a BCC lattice. Contour plots are shown at $\varepsilon_z = 0$, $-0.1$, and $-0.2$.

described in [47]. When negative incremental stiffness (softening) occurs in conventional materials, loss of ellipticity typically leads to localized deformation and thus failure as $v < -1$ is not satisfied. Examples include the shear bands in plastic materials [48-51], collapse bands in cellular materials [52,53], and kink bands in composites [54,55]. Moreover, lumped negative stiffness elements are used to produce high damping composites [3,16] because the negative stiffness materials with multiple elements are not stable even under displacement constraint. Thereby, how to avoid the loss of ellipticity when negative $E$ persists is a fundamental question in preventing material failure as well as realistic applications of negative stiffness materials. Mathematically, an isotropic material with $E < 0$ and $v < -1$ satisfies strong ellipticity thus can prevent the formation of localized bands. A cubic material with $E < 0$, $v < -1$, and $E/(1-2v) + 4G > 0$ also has a great chance of satisfying strong ellipticity [47]. Recently, a 2D metamaterial is demonstrated to show $E < 0$ and $v < -1$ [56], a 3D counterpart is yet to be found. In addition, the above stability analysis also suggests that there's no lower bound of $E$ and $v$ in a constrained material. So, whether there's a mechanism that can have $E$ and $v$ approach negative infinite is another intriguing question.

Here, we develop a 3D metamaterial with negative Poisson's ratio and negative stiffness/bulk modulus by harnessing the snap through instability. Combing parametric studies and theoretical analysis, we show the material can be rationally designed to have negative Poisson's ratio in omnidirectional loads. The Riks method is subsequently used to explore the snap through instability at relative large deformations and the existence of negative stiffness/ compressibility is uncovered. Interestingly, incremental stiffness and Poisson's ratio exhibit singularity when "snap back" is observed in the equilibrium path. The deformation pattern and stress-strain evolution are further analyzed to reveal the underlying mechanism.

The proposed metamaterial adopts the auxetic network concept proposed in [46], where each truss has a large tangential to normal stiffness ratio, analogy to a "sleeved slider unit" [Fig. 1(a)]. We start by finding such "sleeved slider" elements with negative incremental stiffness. Spherical and cylindrical shells are selected as candidates since they can be easily dented but difficult to shear. Fig. 1(b) plots the deformation of several thin shell elements under compression. The two cylinders in contact element (—) undergoes significant lateral expansion thus unqualified. The two spherical shells in contact element (—) presents negligible lateral expansion but no negative post-buckling stiffness. The post-buckling stress rises because the contact

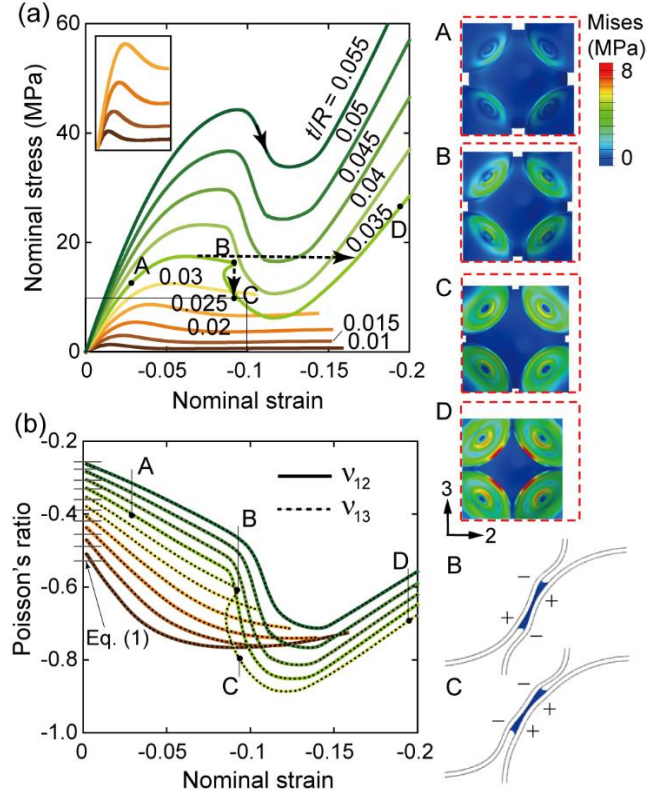

FIG. 2 (a) Nominal stress-strain curves under compression in [100] direction with $r_b/R = 0.2$, $E_b/E_s = 0.04$ and $t/R = 0.01 \sim 0.055$. The vertical and horizontal dashed arrows highlight the path under displacement control and load control, respectively. (b) Evolution of the Poisson's ratio as a function of nominal strain for $t/R = 0.01 \sim 0.055$. The horizontal thin lines present the results of Eq. (1). The stress contours at points A-D are shown on the upper right. Bottom right shows the cross section at (110) plane, "+" and "−" mark the rises and dents of the shells.

area increases, which can be tuned to decrease by introducing a binder between the spherical shells. As depicted by — in Fig. 1(b), the binder-shell elements indent under compression exhibiting a negative incremental stiffness with negligible lateral expansion. Therefore, they are arranged periodically to produce metamaterials. Fig. 1(c) presents the deformation of one resultant metamaterial under uniaxial compression, where negative Poisson's ratio is qualitatively observable. Note that the sphere bulge slightly but the Poisson's ratio is based on measurements of the outer surfaces of the multi sphere specimen. Its negative stiffness will be presented later.

To find the design criteria of the metamaterial, a numerical parametric study is first performed. The discussion is limited to body centered cubic (BCC) lattice for simplicity [Fig. 1(c)]. The numerical simulations were conducted on unit cells with periodic boundary conditions [57,58] utilizing finite element code ABAQUS. The binders are discretized using solid C3D8 elements. The spherical shells are discretized by shell elements and the outer surface of the shells are connected to the bind by offsetting the midplane of shell elements. The accuracy of the mesh was ascertained through a mesh refinement study. The equilibrium path of metamaterials with varying relative shell thickness $t/R$, relative binder radius $r_b/R$, and stiffness ratio $E_b/E_s$ ($h/R$ is fixed at 0.02, and $E_s$ fixed at 70 GPa) under uniaxial compression are calculated using the Riks method [59]. The definition of $t$, $h$, $R$ and $r_b$ are shown in Fig. 1 and subscript "$b$" and "$s$" refer to the binder and shell, respectively.

The simulation results are summarized in Fig. 2 for materials with $h/R = 0.02$, $E_b/E_s = 0.04$, $r_b/R = 0.2$ and varying $t/R$. As expected, the proposed material inherits negative incremental stiffness from the structural elements. Because of cubic symmetry, Poisson's ratio $v_{12}$ overlaps with $v_{13}$ [Fig. 2(b)], as verified by the symmetric deformation pictured in the stress contours. Interestingly, snap back is found for metamaterial with $t/R = 0.035$. The actual path under displacement control and load control will follow the vertical dashed arrow and horizontal dashed arrow, respectively. Under displacement-controlled loading, the force drops abruptly from point B to point C, corresponding to the break of deformation symmetry induced by snap-through instability (Fig. 2 bottom right).

To uncover the underlying mechanism, we start with the mechanical properties at small deformations. Fig. 3(a) and (b) show the Poisson's ratio for $t/R = 0.01 \sim 0.055$, $r_b/R = 0.05 \sim 0.3$, and $E_b/E_s = 0.04 \sim 25$. The Poisson's ratio in this design space is found in the range of $-0.27$ to $-0.77$. To give a theoretical prediction of Poisson's ratio, the following equation is derived based on micromechanics analysis [47],

$$v_{12} = (1 - \lambda)/(2 + \lambda), \qquad (1)$$

where $\lambda$ is the tangential to normal stiffness of the structural element. Results of Eq. (1) is plotted as horizontal lines in Fig. 2(b), which is consistent with the numerical result. Moreover, the simulation results indicate that the normalized effective stiffness scales linearly with $t/R$ and $r_b/R$ (for $r_b/R > 0.1$) by equation

$$E/E_s = C_1 t/R (r_b/R + C_2). \qquad (2)$$

A comparison of Eq. (2) and the numerical results are provided in [47], showing that $C_1 = 1.08$ and $C_2 = 0.0167$ gives an accurate prediction. However, Eq. (1) and (2) only present the material's behavior under [100] direction compression. To obtain the material response in an arbitrary loading direction, simple shear simulations are first performed to calculate $G$. The compliance matrix, $S_{ij}$, is then fully defined by $E$, $v$, and $G$. The compliance matrix, $S'_{ij}$, in an arbitrary new coordinate system defined by Miller indices [$hkl$] and angle of rotation $\theta$ can then be calculated by coordinate transformations [60,61]. As derived in [47], each component of the compliant matrix follows

$$S'_{ij} = S_{ij} + A \cdot F_{ij}([hkl], \theta), \qquad (3)$$

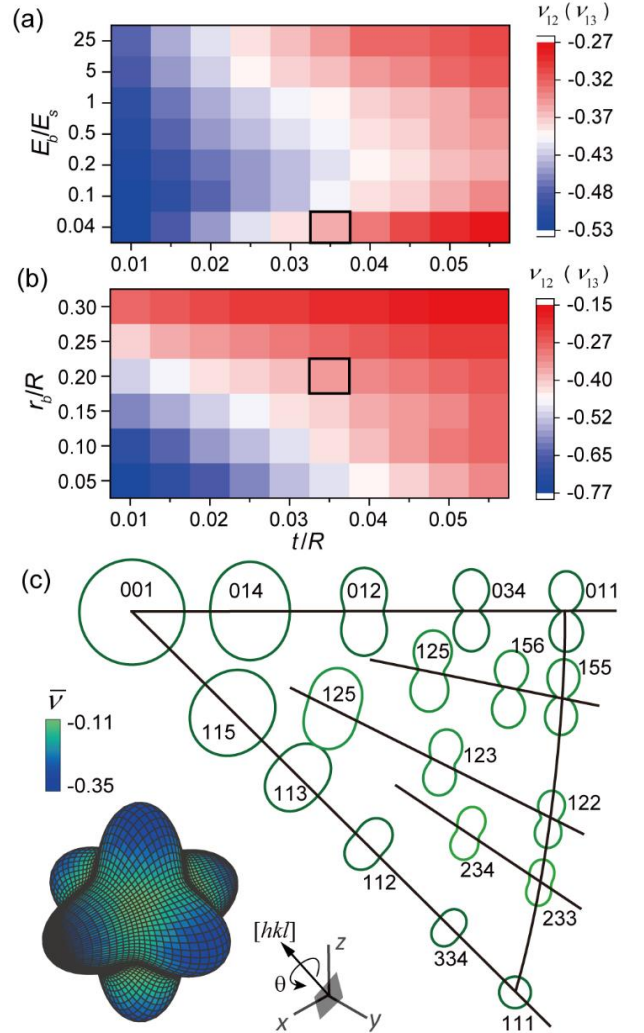

FIG. 3 Results at small deformations. (a-b) Poisson's ratio in the design space of $E_b/E_s = 0.04 \sim 25$, $r_b/R = 0.05 \sim 0.3$, and $t/R = 0.01 \sim 0.55$. (c) The Poisson's ratio on different planes ($hkl$) superposed on the standard triangle of the stereographic projection. The graph is plotted for the metamaterial marked by black rectangles in (a) and (b). The normal of the plane is parallel to the direction of load. The locus at [001] has a radius of $-v_{12} = 0.353$, all other polar plots are drawn to scale. Bottom left plots the average Poisson's ratio $\bar{v}$ in a 3D representation surface.

where $A = (S_{11} - S_{12} - S_{44}/2)$ and expressions of $F_{ij}$ are functions provided in [47]. Poisson's ratio with respect to load direction $[hkl]$ and transverse direction defined by $\theta$ can be calculated by $\nu_{[hkl],\theta} = -S_{12}'/S_{11}'$ (illustrated in Fig. S7). Since Poisson's ratio depends on both $[hkl]$ and $\theta$, visualizing $\nu_{[hkl],\theta}$ is not trivial. Fig. 3(c) uses a stereographic projection [62] to present the Poisson's ratio of the metamaterial with $t/R = 0.035$, $r_b/R = 0.2$, and $E_b/E_s = 0.04$. Base on this plot, it is evident that when the loading direction changes from [001] to [011], the dependency of $\nu_{[hkl],\theta}$ on $\theta$ increases. Generally, the Poisson's ratio on an arbitrary plane exhibits a two-fold symmetry for a cubic material, except that in the [100] and [111] direction. Specifically, if compressed in [100] or [111] direction, the material will contract isotropically in the lateral directions. By contrast, elliptical or peanut shaped angle dependency of Poisson ratio is exhibited when loading the metamaterial in other directions. Furthermore, the average Poisson's ratio in a certain loading direction, $\bar{\nu}_{[hkl]}$, is calculated by integrating over $\theta$, which is plotted as a 3D representation surface [63] in Fig. 3(c). By searching the solution space numerically, it is found $-0.35 < \bar{\nu} < -0.11$, suggesting that the metamaterial exhibits an omnidirectional negative Poisson's ratio. Note that this is not a common feature for the proposed metamaterial, whether a material exhibit omnidirectional negative Poisson's ratio should be checked by evaluating the maximum $\nu$.

Next, we focus on the material's post-buckling behavior. To characterize the mechanical performance at relatively large deformations, the incremental stiffness, $E^{inc} = d\sigma_{11}/d\varepsilon_{11}$ and incremental Poisson's ratio, $\nu^{inc} = -d\varepsilon_{22}/d\varepsilon_{11}$, are calculated. As shown in Fig. 4 (a) and (b), dips of $E^{inc}$ and $\nu^{inc}$ are prominent. Interestingly, when snap back takes place ($t/R = 0.035$), both $E^{inc}$ and $\nu^{inc}$ show singularity approaching negative infinity. To uncover the mechanism of the "anomalous" decrease of $E^{inc}$ and $\nu^{inc}$, the evolution of transverse strains and deformations of the connected sphere pairs are plotted in Fig. 4(c) and (d), respectively. For $t/R = 0.035$, Fig. 4 show that when $\varepsilon_{11}$ reaches a critical strain, an infinite small decrease of $\varepsilon_{11}$ will lead to a finite reduction of stress and transverse strain. Therefore, the incremental parameter $E^{inc}$ and $\nu^{inc}$ approach negative infinity. The deformation fields reveal that snap back is accompanied by a change of the sphere's deformation pattern. To verify that the observed anomalies are caused by pattern transformation of the spherical shells, we compare two metamaterials with ($t/R = 0.04$) and without ($t/R = 0.03$) pattern transformations. Numerical results show that for $t/R = 0.03$, the two adjacent spheres maintain a consistent symmetric deformation pattern [Fig. 4(d)]. As such, both $E^{inc}$ and $\nu^{inc}$ vary smoothly, exhibiting no anomalies. In contrast, for $t/R = 0.04$, the deformation of the adjacent spheres change from a symmetric pattern to a non-symmetric pattern, which is presented in Fig. 4(d) and [47]. Moreover, the range of applied strain where pattern transformation takes place (between point B and point C) is consistent with the range of negative $E^{inc}$ (shaded area) and a fast decrease of transverse strain $\varepsilon_{22}$ and $\varepsilon_{33}$. To give a quantitative view of the pattern transformation, the evolution of the strain energies of two adjacent spheres is plotted in the inset of Fig. 4(c), showing that the strain energy bifurcates as pattern transformation is initiated. These results conclude that the anomalies in $E^{inc}$ and $\nu^{inc}$ result from the break of deformation symmetry, which causes an abrupt stress drop and an abrupt lateral contraction.

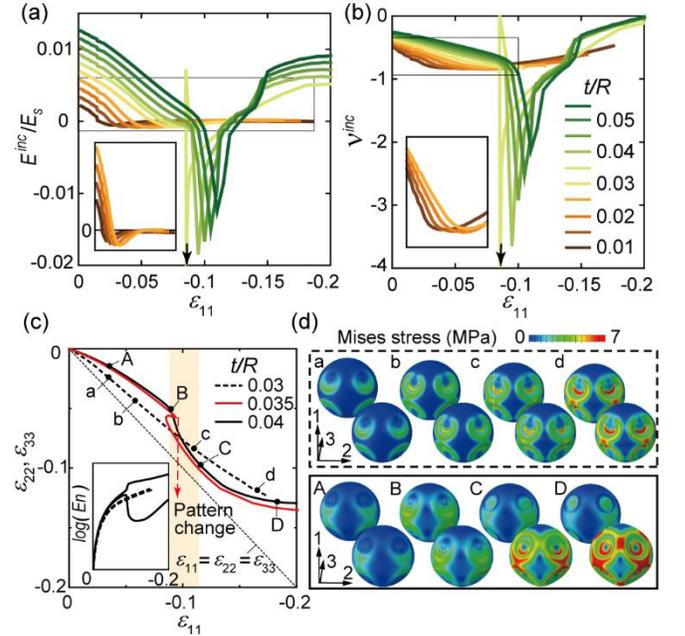

FIG. 4 (a) Incremental stiffness and (b) Incremental Poisson's ratio at relatively large deformations. Singularity emerges when snap back is observed in the equilibrium path. The arrows highlight the singularity approaching negative infinity. The insets show rescaled curves for easy viewing. (c) Evolution of transverse strains versus applied strain $\varepsilon_{11}$. The shaded area highlights the region of negative $E^{inc}$ for $t/R = 0.04$. Inset shows the evolution of strain energy ($En$) of two adjacent spheres. (d) The deformation of contacted spherical shells is plotted for points a-d and A-D.

So far, the numerical simulations are based on unit cells. It is acknowledged that when pattern transformation is triggered, the microscopic bifurcation may have a longer periodic length than the unit cell. To verify that the unit cell suffices to capture the microscopic instability, we further perform simulations on super unit cells with 2×2×2, 1×1×4, 2×2×4, and 3×3×3 unit cells referring to [64]. The pattern transformation of all the super unit cells is found to be the same as the one unit cell result as shown in Fig. 4(d) [47]. In addition, since the metamaterial presents snap through type instability, adding imperfection is not necessary to initiate buckling. For a realistic material, however, imperfection is inevitable. To predict the effect of imperfection, eigenmodes are artificially introduced to the material as imperfections. Results show that imperfections will reduce the value of negative stiffness and negative Poisson's ratio, nevertheless, the double negative indices still exist [47].

As discussed previously, both $E^{inc}$ and $\nu^{inc}$ approach negative infinite when pattern transformation takes place instantaneously (snap back is presented). This is a direct numerical verification to the theoretical prediction that $E \to -\infty$ and $\nu \to -\infty$ are allowed physically. In contrast, when pattern transformation takes place in a range of applied strains, finite values of $E^{inc} < 0$ and $\nu^{inc} < -1$ can be obtained. In previous studies, negative stiffness materials have positive Poisson's ratio, which does not satisfy strong ellipticity. Here, after checking the wave speed in all possible wave propagation directions and wave polarizations for the metamaterial with $r_b/R = 0.2$, $E_b/E_s = 0.04$ and $t/R = 0.04$, we found it satisfies the strong ellipticity condition [47]. A metamaterial with $E^{inc} < 0$ and satisfies strong ellipticity is of great application importance because then a large block of the metamaterial (containing many unit cells) can be used. Also note that the incremental stiffness and Poisson's ratio become the "applicable" stiffness and Poisson's ratio when prestressed to the corresponding strain. Thereby, prestressed metamaterials can be used to create composites with exceptional high damping and high stiffness as reported in [3,11,16,17]. Furthermore, bulk modulus can be calculated by $K = E/3(1-2\nu)$. The proposed metamaterial with negative $E$ and $\nu$ also leads to a negative bulk modulus. This is known as negative compressive metamaterial and has potential applications like actuators, force amplifiers, and micromechanical controls [65].

In summary, we demonstrate a triple negative cubic metamaterial that can satisfy the strong ellipticity condition. Without prestress (at small strains), the material exhibits negative Poisson's ratio, which can be rationally tuned by the geometric parameters. Evaluating the Poisson's ratio in arbitrary loading directions reveals an omnidirectional negative Poisson's ratio. When prestressed to specific strain levels, the metamaterial presents both negative stiffness, negative bulk modulus, and negative Poisson's ratio. Moreover, $E^{inc}$ and $\nu^{inc}$ approach negative infinity when snap back occurs. Although the metamaterial possesses $E^{inc} < 0$, it satisfies the strong ellipticity condition, thus opens new possibility to prevent deformation localization and enable the application of homogeneous negative stiffness/compressibility materials. Importantly, recent progress in self-assembling techniques [66-68] makes fabrication of the metamaterial at nanoscale possible, which will facilitate new applications of multi-negative index materials.

A sincere thanks goes to Professor Roderic Lakes for his insightful comments and valuable suggestions during the course of this research. This research was supported by the the National Science Foundation (CMMI-1462270).

––––––––––

[1] K. Bertoldi, V. Vitelli, J. Christensen, and M. van Hecke, Nat. Rev. Mater. **2**, natrevmats201766 (2017).

[2] X. Yu, J. Zhou, H. Liang, Z. Jiang, and L. Wu, Prog. Mater Sci. **94**, 114 (2018).

[3] R. S. Lakes, T. Lee, A. Bersie, and Y. Wang, Nature **410**, 565 (2001).

[4] R. Lakes, Science **235**, 1038 (1987).

[5] C. Kane and T. Lubensky, Nat. Phys. **10**, 39 (2014).

[6] G. Milton, *The theory of composites* (Cambridge University Press, Cambridge, 2002).

[7] R. H. Baughman, S. Stafström, C. Cui, and S. O. Dantas, Science **279**, 1522 (1998).

[8] C. Coulais, D. Sounas, and A. Alù, Nature **542**, 461 (2017).

[9] J. Cherfas, Science **247**, 630 (1990).

[10] G. W. Milton, J. Mech. Phys. Solids. **40**, 1105 (1992).

[11] R. Lakes and W. Drugan, J. Mech. Phys. Solids. **50**, 979 (2002).

[12] S. Xinchun and R. S. Lakes, Physica status solidi (b) **244**, 1008 (2007).

[13] S. Mohsenizadeh, R. Alipour, M. S. Rad, A. F. Nejad, and Z. Ahmad, Materials & Design **88**, 258 (2015).

[14] S. Hou, T. Li, Z. Jia, and L. Wang, Materials & Design **160**, 1305 (2018).

[15] S. Shan, S. H. Kang, J. R. Raney, P. Wang, L. Fang, F. Candido, J. A. Lewis, and K. Bertoldi, Adv. Mater. **27**, 4296 (2015).

[16] R. Lakes, Philos. Mag. Lett. **81**, 95 (2001).

[17] T. Jaglinski, D. Kochmann, D. Stone, and R. Lakes, Science **315**, 620 (2007).

[18] B. Xu, F. Arias, S. T. Brittain, X.-M. Zhao, B. Grzybowski, S. Torquato, and G. M. Whitesides, Adv. Mater. **11**, 1186 (1999).

[19] C. Lees, J. F. Vincent, and J. E. Hillerton, Bio-Med. Mater. Eng. **1**, 19 (1991).

[20] D. Kang, M. P. Mahajan, S. Zhang, R. G. Petschek, C. Rosenblatt, C. He, P. Liu, and A. Griffin, Phys. Rev. E **60**, 4980 (1999).

[21] R. H. Baughman, J. M. Shacklette, A. A. Zakhidov, and S. Stafström, Nature **392**, 362 (1998).

[22] F. Milstein and K. Huang, Phys. Rev. B **19**, 2030 (1979).

[23] Z. Zhang, H. Davis, and D. Kroll, Phys. Rev. E **53**, 1422 (1996).

[24] M. Bowick, A. Travesset, G. Thorleifsson, and A. Cacciuto, Phys. Rev. Lett. **87**, 148103 (2001).

[25] J.-W. Jiang and H. S. Park, Nano Lett. **16**, 2657 (2016).

[26] I. Shufrin, E. Pasternak, and A. V. Dyskin, Int. J. Solids Struct. **54**, 192 (2015).

[27] M. Schenk and S. D. Guest, Proc. Natl. Acad. Sci. **110**, 3276 (2013).

[28] H. Yasuda and J. Yang, Phys. Rev. Lett. **114**, 185502 (2015).

[29] T. Li, X. Hu, Y. Chen, and L. Wang, Sci. Rep. **7**, 8949 (2017).

[30] K. Bertoldi, P. M. Reis, S. Willshaw, and T. Mullin, Adv. Mater. **22**, 361 (2010).

[31] S. Babaee, J. Shim, J. C. Weaver, E. R. Chen, N. Patel, and K. Bertoldi, Adv. Mater. **25**, 5044 (2013).

[32] Y. Chen, T. Li, F. Scarpa, and L. Wang, Phys. Rev. Appl. **7**, 024012 (2017).

[33] Y. Cho *et al.*, Proc. Natl. Acad. Sci. **111**, 17390 (2014).

[34] C. Coulais, A. Sabbadini, F. Vink, and M. van Hecke, Nature **561**, 512 (2018).

[35] D. Prall and R. Lakes, Int. J. Mech. Sci. **39**, 305 (1997).

[36] A. Pozniak and K. Wojciechowski, physica status solidi (b) **251**, 367 (2014).

[37] Y. Jiang and Y. Li, Adv. Eng. Mater. **19**, 1600609 (2017).

[38] T. Bückmann, R. Schittny, M. Thiel, M. Kadic, G. W. Milton, and M. Wegener, New J. Phys. **16**, 033032 (2014).

[39] F.-K. Chang and K.-Y. Chang, J. Compos. Mater. **21**, 834 (1987).

[40] Z. Liu, X. Zhang, Y. Mao, Y. Zhu, Z. Yang, C. T. Chan, and P. Sheng, Science **289**, 1734 (2000).

[41] N. Fang, D. Xi, J. Xu, M. Ambati, W. Srituravanich, C. Sun, and X. Zhang, Nat. Mater. **5**, 452 (2006).

[42] E. Salje, Ferroelectrics **104**, 111 (1990).

[43] J. Qiu, J. H. Lang, and A. H. Slocum, Journal of microelectromechanical systems **13**, 137 (2004).

[44] S. P. Timoshenko and J. Goodier, *Theory of elasticity* (McGraw-Hill Publishing Company, 1970).

[45] J. K. Knowles and E. Sternberg, Journal of Elasticity **8**, 329 (1978).

[46] L. Rothenburg, A. A. Berlin, and R. J. Bathurst, Nature **354**, 470 (1991).

[47] Supplemantary material available online, for additional experimental, simulation and theoretical details.

[48] D. Bigoni, *Nonlinear solid mechanics: bifurcation theory and material instability* (Cambridge University Press, 2012).

[49] R. Hill, J. Mech. Phys. Solids. **10**, 1 (1962).

[50] J. R. Rice, in *14th International Congress of theoretical and applied mechanics* (OSTI.gov, Netherlands, 1976).

[51] S. Harren, H. Deve, and R. Asaro, Acta Metall. **36**, 2435 (1988).

[52] S. D. Papka and S. Kyriakides, Int. J. Solids Struct. **35**, 239 (1998).

[53] S. D. Papka and S. Kyriakides, Acta Mater. **46**, 2765 (1998).

[54] R. Gutkin, S. Pinho, P. Robinson, and P. Curtis, Compos. Sci. Technol. **70**, 1223 (2010).

[55] M. Harraez, A. C. Bergan, C. Gonzalez, and C. S. Lopes, (2018).

[56] T. A. Hewage, K. L. Alderson, A. Alderson, and F. Scarpa, Adv. Mater. **28**, 10323 (2016).

[57] M. Danielsson, D. Parks, and M. Boyce, J. Mech. Phys. Solids. **50**, 351 (2002).

[58] L. Wang, M. C. Boyce, C. Y. Wen, and E. L. Thomas, Adv. Funct. Mater. **19**, 1343 (2009).

[59] E. Riks, Int. J. Solids Struct. **15**, 529 (1979).

[60] J. Turley and G. Sines, J. Phys. D: Appl. Phys. **4**, 1731 (1971).

[61] T. Thomas, Proc. Natl. Acad. Sci. **55**, 235 (1966).

[62] K. M. Knowles and P. R. Howie, Journal of Elasticity **120**, 87 (2015).

[63] J. F. Nye, *Physical properties of crystals: their representation by tensors and matrices* (Oxford university press, 1985).

[64] M. Laroussi, K. Sab, and A. Alaoui, Int. J. Solids Struct. **39**, 3599 (2002).

[65] Z. G. Nicolaou and A. E. Motter, Nat. Mater. **11**, 608 (2012).

[66] B. Hatton, L. Mishchenko, S. Davis, K. H. Sandhage, and J. Aizenberg, Proc. Natl. Acad. Sci. **107**, 10354 (2010).

[67] A. Dong, X. Ye, J. Chen, and C. B. Murray, Nano Lett. **11**, 1804 (2011).

[68] Z. Chen, P. Zhan, Z. Wang, J. Zhang, W. Zhang, N. Ming, C. T. Chan, and P. Sheng, Adv. Mater. **16**, 417 (2004).

**Finding a negative stiffness material that satisfies the strong ellipticity condition**

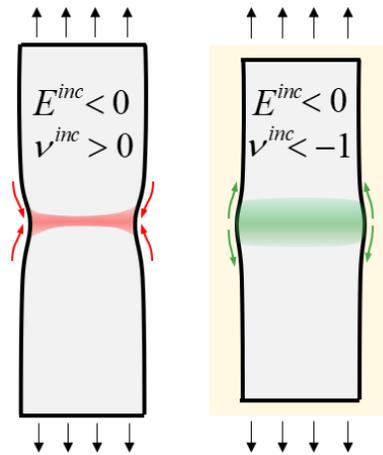

**TOC Figure.** Traditional material with negative incremental stiffness will form localized deformation bands, which often leads to failure. By contrast, a deformation band can be delocalized in a double negative metamaterial, providing a mechanism to prevent premature failure.

# Supplementary Material for
## 3D mechanical metamaterials with negative stiffness and negative Poisson's ratio approaching negative infinity

Zian Jia[1], and Lifeng Wang[1, *]

*[1]Department of Mechanical Engineering, State University of New York at Stony Brook, Stony Brook, New York 11794, USA*

## Numerical simulations

The geometry of the metamaterial is modelled referring to foam materials fabricated by sintering and self-binding [1-3] [Fig. S1(a)]. Specifically, the binders are constructed by revolving arcs that tangentially connect two adjacent spheres. The binders and shells are assumed to have different materials.

The mechanical performance of the metamaterial is calculated on unit cells with periodic boundary conditions [4,5] [Fig. S1(b)]. A periodic material subjected to a macroscopic deformation gradient $\mathbf{F}$ should satisfy the following equation [4],

$$\mathbf{u}(B) - \mathbf{u}(A) = (\mathbf{F} - \mathbf{I})\{\mathbf{X}(B) - \mathbf{X}(A)\} = \mathbf{H}\{\mathbf{X}(B) - \mathbf{X}(A)\}, \tag{S1}$$

where $A$ and $B$ are a pair of points located on the periodic boundaries. $\mathbf{u}$ denotes the displacement, $\mathbf{X}$ denotes the position in the reference configuration, and $\mathbf{H} = \mathbf{F} - \mathbf{I}$ is the macroscopic displacement gradient tensor. For a 3D structure, the displacement gradient is a 3×3 matrix

$$\mathbf{H} = \begin{bmatrix} H_{11} & H_{12} & H_{13} \\ H_{21} & H_{22} & H_{23} \\ H_{31} & H_{32} & H_{33} \end{bmatrix} = \begin{bmatrix} F_{11}-1 & F_{12} & F_{13} \\ F_{21} & F_{22}-1 & F_{23} \\ F_{31} & F_{32} & F_{33}-1 \end{bmatrix}. \tag{S2}$$

In finite element implementation (commercial software ABAQUS), Eq. (S1) is implement by equation constraints and $\mathbf{H}$ is assigned by the displacement components of three reference nodes. In this study, uniaxial compression of the unit cell in [100] direction is calculated by assigning $H_{11} = \varepsilon_{11}$. The transverse strains are subsequently calculated from the resultant components $H_{22}$ and $H_{33}$. The macroscopic first Piola-Kirchoff stress tensor is then extracted through virtual work consideration [4]. In addition, shear modulus $G$ is calculated by applying a simple shear to the unit cell with $H_{12} = \gamma$ and all other components being zero.

The snap through and snap back buckling behavior of the metamaterial is captured by the Riks method [6]. The binders are discretized by C3D8 element and shells are discretized using shell element S3 and S4, which uses thick shell theory or Kirchhoff thin shell theory based on the shell thickness. The *offset command is used to offset the midplane of shell elements such that the outer surface of the spherical shell connects to the binder.

Using the described method, the stress-strain relationship in [100] direction under uniaxial compression is calculated, as shown in Fig. 2 in the main text. To highlight the role of geometric nonlinearity, linear elastic materials are assumed in the simulation. Specifically, the material properties of the shell are $E_s = 70$ $GPa$ and $\nu_s = 0.3$. The Poisson's ratio of the binder is $\nu_b = 0.4$, and its modulus $E_b$ is defined by assigning $E_b/E_s$.

Fig. S2 summarizes the lateral expansions of the structural elements under compression. Results show that the cylinder shells present two orders greater bulging effect than the spherical shells. The deformation of periodically arranged 2D cylinders is shown in Fig. S3, which has a positive Poisson's ratio of 0.66 because of the significant lateral expansion (bulging effect). To verify that the conventional BCC unit cell successfully captures the buckling pattern, simulations on super unit cells with 2×2×2, 2×2×8, and 3×3×3 conventional unit cells are also performed. The comparison in Fig. S8 shows that all simulations exhibit the same deformation pattern, verifying the use of the conventional unit cell.

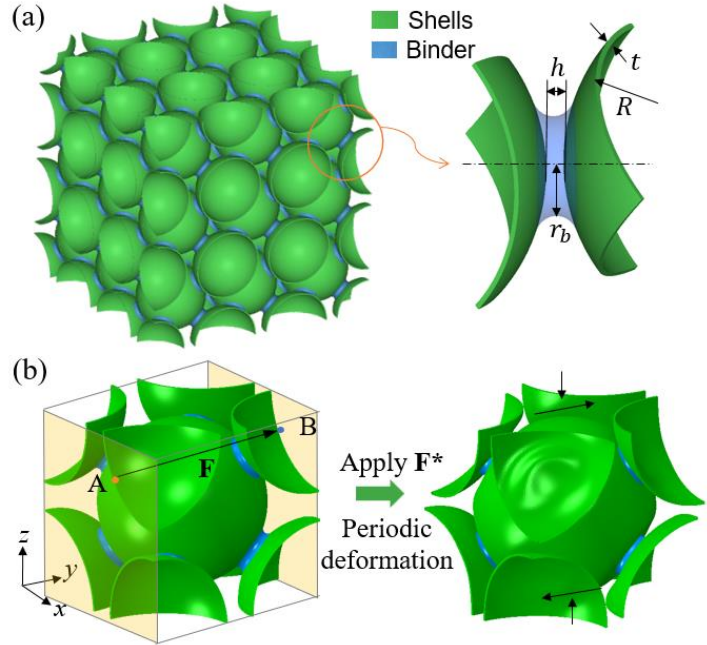

FIG. S1 (a) Geometry of the periodic 3D metamaterial. (b) Conventional unit cell and the periodic boundary condition. Left shows the periodic constraints (defined by $\mathbf{F}$) applied on a pair of points on the boundary. Right demonstrates a periodic deformation with $\mathbf{F}^*_{33} = 0.9$, $\mathbf{F}^*_{23} = 0.1$, all other components of $\mathbf{F}^*$ are zero.

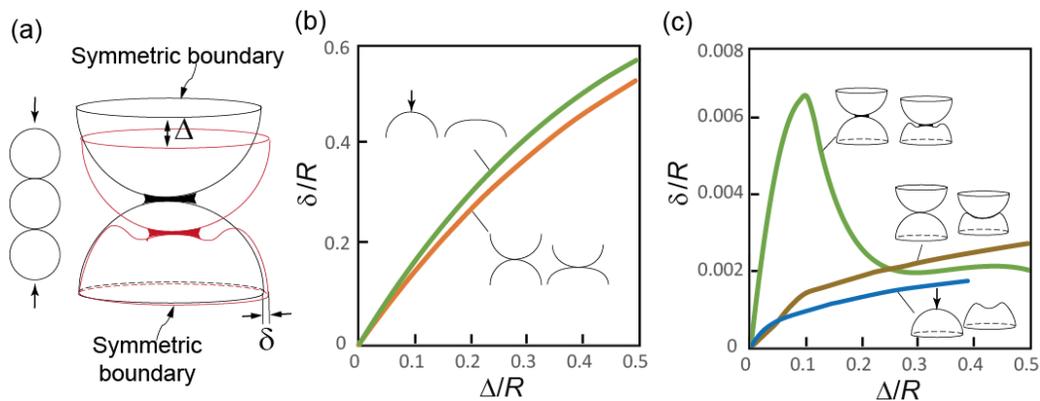

FIG. S2 Lateral deformations of the structural elements under compression. (a) Numerical model, definition of the applied displacement $\Delta$, and the lateral expansion $\delta$. (b) and (c) show the $\delta/R$ vs $\Delta/R$ relation of cylindrical shell elements and spherical shell elements respectively. The spherical elements show a significantly smaller "bulging" effect.

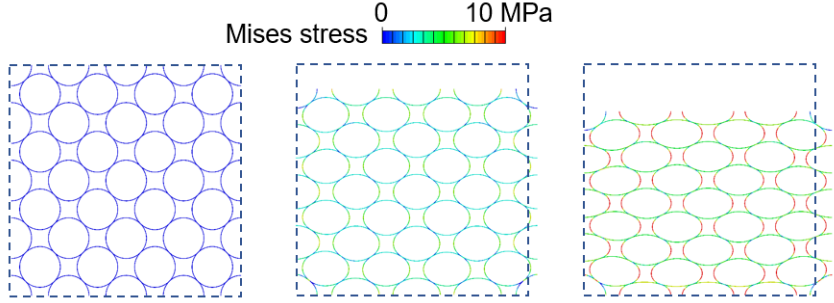

FIG. S3 Deformation of a 2D periodically arranged cylindrical shells under compression. The 2D counterpart to the proposed 3D metamaterial exhibits a positive Poisson's ratio of 0.66, because each cylinder "bulges" instead of "indents".

## Micromechanics analysis

In order to derive a theoretical prediction of Poisson's ratio under small deformation, we perform a micromechanics analysis on the unit cell (Fig. S4). The analysis is based on the force equilibrium in the (100) plane. Assume the force applied on each binder-shell element in the direction [100] is $\mathbf{F}$. The force components in the normal and tangent directions are $\mathbf{F}_n$ and $\mathbf{F}_t$, respectively [Fig. S4(b)]. The corresponding displacements are $\delta_n$ and $\delta_t$. Note the normal and tangential stiffness of the structural element as $k_n$ and $k_t$, the corresponding displacements become

$$\delta_n = F_n / K_n , \; \delta_t = F_t / K_t . \tag{S3}$$

The resultant deformations in [100] and [011] directions contributed by $\delta_n$ and $\delta_t$ can be written as

$$\delta_{[100]} = \delta_n \sin \theta + \delta_t \cos \theta , \tag{S4}$$

$$\delta_{[011]} = \delta_n \cos \theta - \delta_t \sin \theta . \tag{S5}$$

At small deformations, the displacements in the [010] and [001] directions are related to the displacement in the [011] direction by

$$\delta_{[010]} = \delta_{[001]} = \delta_{[011]} / \sqrt{2} . \tag{S6}$$

Combining the Eq. (S3-S6) with relation $F_n = F \sin \theta$ and $F_t = F \cos \theta$, we have

$$\nu_{12} = \frac{\delta_{[010]}}{\delta_{[100]}} = \frac{\sin \theta \cos \theta (1 - k_t / k_n)}{\sqrt{2}(\cos^2 \theta + \sin^2 \theta \cdot k_t / k_n)} . \tag{S7}$$

In a BCC lattice, $\sin \theta = 1 / \sqrt{3}$ , $\cos \theta = \sqrt{2} / \sqrt{3}$ . Plug into Eq. (S7) and denote $k_t / k_n$ as $\lambda$, we derive Eq. (1) in the main text

$$\nu_{12} = \nu_{13} = \frac{1 - \lambda}{2 + \lambda} . \tag{S8}$$

This equation indicates that structural elements with $\lambda > 1$ will lead to negative Poisson's ratio, which is consistent with results in [7]. An energy based analysis gives the same result [8]. With this equation, Poisson's ratio can be predicted by calculating the normal and tangential stiffnesses of the structural elements numerically.

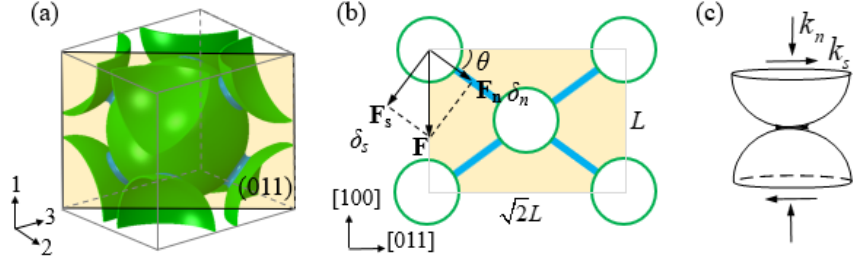

FIG. S4 Schematics for deriving Eq. (1). (a) The light-yellow color highlights the (011) plane used for analysis. The unit cell is under uniaxial compression in direction 1. (b) **F** denotes the resultant force on each binder-shell element in direction [100]. $F_n$ and $F_t$ denote the force components in the normal and tangent directions, $\delta_n$ and $\delta_t$ denote the corresponding displacements. The spheres and binders are not drawn to scale for easy viewing. (c) Definition of the normal and tangential stiffness of the binder-shell structural element.

Moreover, the formula for predicting Young's modulus is obtained by fitting the numerical results. For materials with $E_b/E_s = 0.04$ and $h/R = 0.02$ the normalized effective stiffness is found to scale linearly with $t/R$ and $r_b/R$ by equation

$$E/E_s = 1.08 \frac{t}{R} \left( \frac{r_b}{R} + 0.0167 \right). \tag{S9}$$

A comparison between the prediction of Eq. (2) and numerical simulations are provided in Fig. S5, showing a good agreement.

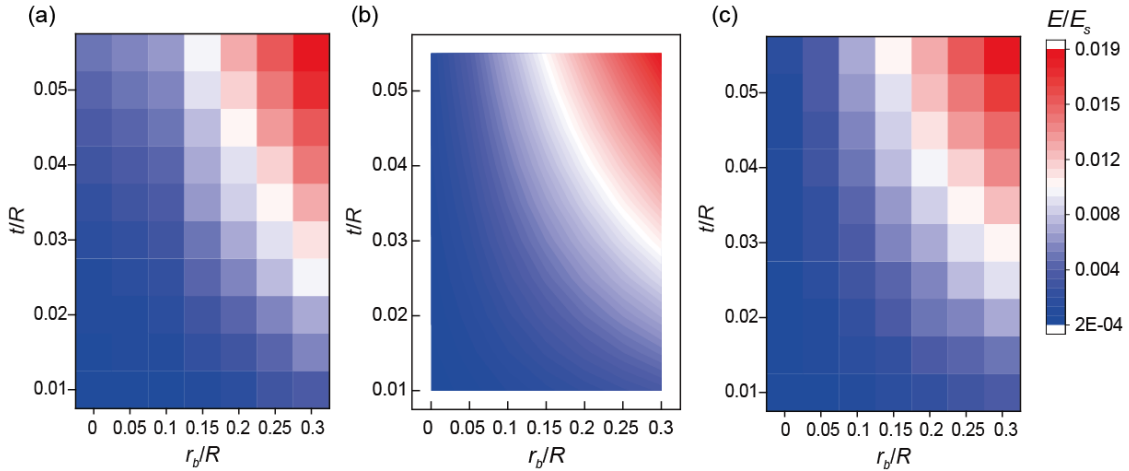

FIG. S5 Comparison of $E/E_s$ between Eq. (2) with $C_1$=1.08 and $C_2$=0.0167 and numerical results. (a) Numerical simulation results. (b) Theoretical prediction plotted in a continuous contour and (c) theoretical prediction in a Mosaic plot.

# Stability analysis

*Unconstrained materials*

In unconstrained materials, the positive definite strain energy condition requires energy densities including $E\varepsilon_{11}^2/2$, $G\gamma_{12}^2/2$, and $K\varepsilon_B^2/2$ to be positive, therefore, $E$, $G$, and $K$ should be positive. For isotropic materials, $G = E/2(1+\nu)$, $K = 3(1-2\nu)/E$, the positive requirement of $E$, $G$, and $K$ derives $-1 < \nu < 0.5$. For cubic materials, by contrast, only the equation $K = 3(1-2\nu)/E$ holds, thus the stability condition is looser, $\nu < 0.5$.

*The strong ellipticity condition for constrained materials*

To ensure that a material is stable under displacement constraint, the strong ellipticity condition must be satisfied [9]. The strong elliptic condition poses the following requirement on the elastic tensor, $C_{ijkl}n_i n_j m_k m_l > 0$, where $n_i$ and $m_i$ are arbitrary unit vectors [10,11]. An alternative form of this requirement is that the acoustic tensor $A_{ik} = C_{ijkl}n_j n_l$ should be positive definite for all possible $n$. Here, the discussions are based on the acoustic tensor.

For an isotropic material, the three wave propagation speeds $c_L^2$, $c_S^2$, and $c_S^2$ are calculated form the eigenvalues of the acoustic tensor, which are direction independent and has the following expressions [12],

$$c_L^2 = \frac{E(1-\nu)}{\rho(1+\nu)(1-2\nu)}, \; c_S^2 = \frac{G}{\rho}. \tag{S10}$$

where the subscripts $L$ and $S$ indicate the longitudinal mode and shear mode respectively. Positive definite $A_{ik}$ thus requires $\frac{E(1-\nu)}{(1+\nu)(1-2\nu)} > 0$ and $G > 0$. Physically, this means the material should have real wave speeds to propagate the deformation to the whole material. Based on Eq. (S10), the material property space that satisfies the strong ellipticity condition can be obtained. The resultant material parameter space is summarized in Table S1, which has three regions [13].

**Table S1**. Material parameters satisfying the strong ellipticity condition for isotropic materials under displacement constraint.

| Region 1, negative $\nu$ | Region 2, triple negative | Region 3, negative $K$ |
|---|---|---|
| $G > 0$ | $G > 0$ | $G > 0$ |
| $E > 0$, $-1 < \nu < 0.5$ | $E < 0$, $\nu < -1$ | $E > 0$, $\nu > 1$ |
| $K > 0$ | $-G/3 < K < 0$ | $-4G/3 < K < -G/3$ |

Next, let's discuss the strong ellipticity condition for cubic materials. The wave propagation in anisotropic media is different from isotropic materials because the elastic wave velocity is direction dependent. Thereby, a rigorous verification of the strong ellipticity condition requires checking that the material has positive wave propagation speed in all wave incident directions and wave polarizations. This is achieved by numerically calculating the eigenvalues of the acoustic tensor, $A_{ik} = C_{ijkl}n_j n_l$, for all wave propagation directions, with $n$ being the wave propagation direction. The propagation direction of an arbitrary wave is defined by the spherical coordinates as $n = [\cos\theta\sin\varphi, \sin\theta\sin\varphi, \cos\varphi]^T$, where $\theta$ is the azimuth angle and $\varphi$ is the zenith angle. The complete range of propagation directions is computed with $\theta$ varying from –180º to 180º and $\varphi$ varying from 0º to 180º. Given $C_{ijkl}$ and $n$, $A_{ik}$ can be calculated and its three eigenvalues correspond to three wave polarizations (one quasi-longitudinal, and two quasi-shear), $\rho c_L^2$, $\rho c_{S1}^2$, and $\rho c_{S2}^2$.

For a cubic symmetric material the stiffness tensor $C_{ijkl}$ is defined by three parameters, $C_{11}$, $C_{12}$, and $C_{44}$. Specifically, $C_{11} = \dfrac{E(1-\nu)}{(1+\nu)(1-2\nu)}$, $C_{12} = \dfrac{E\nu}{(1+\nu)(1-2\nu)}$, and $C_{44} = G$. These parameters are calculated with FEM described in the numerical simulation methods section. In the main text, we've discussed the elliptic condition for the metamaterial with $r_b/R = 0.2$, $E_b/E_s = 0.04$ and $t/R = 0.04$. At compression strain of 0.095, the material presents $E^{inc} = -1194.7$ MPa, $\nu^{inc} = -3.652$, $G^{inc} = 90.04$ MPa ($G$ is evaluated with FEM by first applying $\varepsilon_{11} = -0.095$, then apply a small shear strain). A MATLAB program is developed to realize the described procedure. The flowchart of the program and resultant wave speeds in terms of $\rho c^2$ are presented in Fig. S6. Results proved that the metamaterial presents positive wave speed in all wave propagation directions, therefore, the strong ellipticity condition is verified.

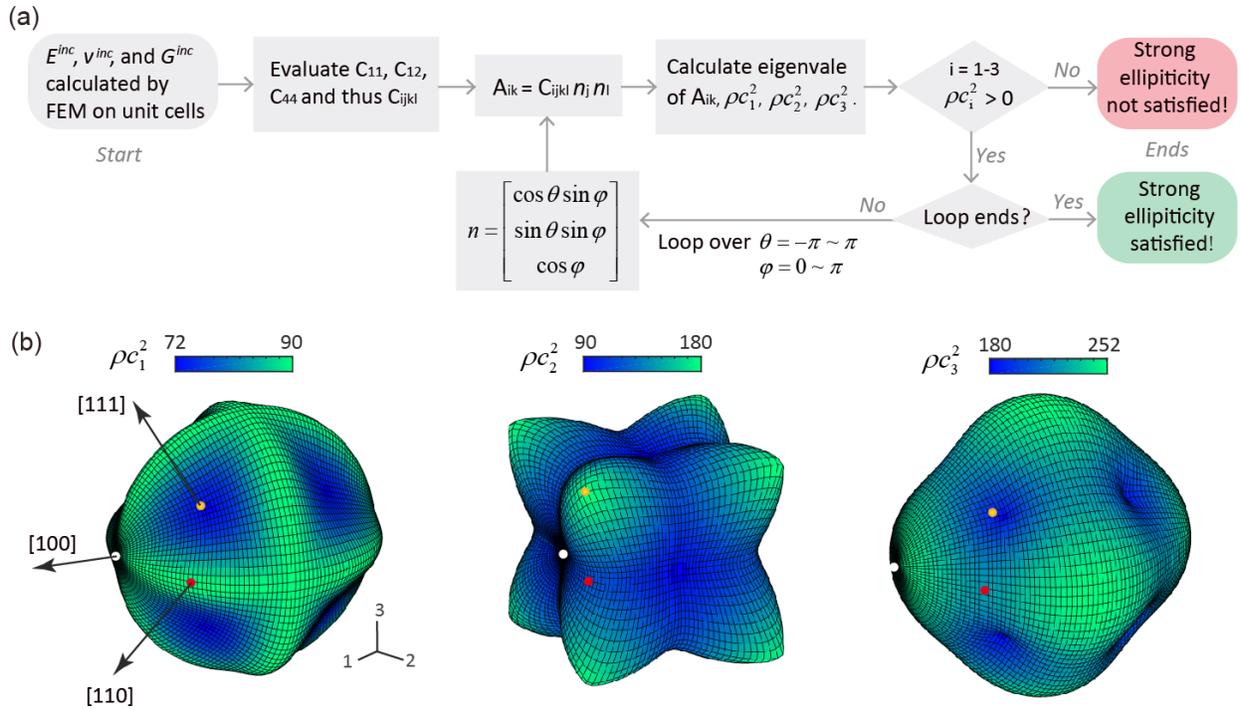

FIG. S6 Checking the strong ellipticity condition for cubic materials. (a) Flowchart showing the procedure used to check the strong ellipticity condition. (b) The wave speeds of a metamaterial with $E^{inc} = -1194.7$ MPa, $\nu^{inc} = -3.652$, $G^{inc} = 90.04$ MPa at strain $\varepsilon_{11} = -0.095$. Results show that this metamaterial presents positive wave speeds in all directions, thus satisfies the strong ellipticity condition. Note that $c_1$, $c_2$, and $c_3$ corresponds the minimum, secondary, and maximum wave speed respectively. Directions with high symmetry are also marked.

The above method requires an exhaustive search on all the wave propagation directions, which is troublesome. A simplified approach is to check the wave speeds in the high symmetric planes, i.e., [100], [110], and [111] direction, where analytical solutions exist. Although not rigorous, this approach provides a necessary condition for strong ellipticity. Moreover, the minimum wave propagation speed (critical point) typically appears on the high symmetric planes as exhibited in Fig. S6. Therefore, an analytical calculation of wave speeds in [100], [110], and [111] directions can serve as a preliminary check of strong ellipticity. For cubic materials, the wave speeds in [100] direction are [14]

$$c_L^2 = \frac{C_{11}}{\rho}, \ c_{S1}^2 = c_{S2}^2 = \frac{C_{44}}{\rho}. \tag{S11}$$

The wave propagation speed condition in this direction is the same as that for isotropic materials. Positive wave speed condition entails $c_L^2 = \frac{E(1-\nu)}{\rho(1+\nu)(1-2\nu)} > 0$ and $c_S^2 = \frac{G}{\rho} > 0$.

Moreover, the wave speeds in [110] directions are [14]

$$c_L^2 = \frac{C_{11} + C_{12} + 2C_{44}}{2\rho}, \ c_{S1}^2 = \frac{C_{11} - C_{12}}{2\rho}, \ c_{S2}^2 = \frac{C_{44}}{\rho}. \tag{S12}$$

And the wave speeds in [111] directions are [14]

$$c_L^2 = \frac{C_{11} + 2C_{12} + 4C_{44}}{3\rho}, \ c_{S1}^2 = c_{S2}^2 = \frac{C_{11} - C_{12} + C_{44}}{3\rho}. \tag{S13}$$

The real wave speed requirements entail the right hand sides of Eq. (S11-S13) to be positive, which can be summarized as $C_{11} > 0$, $C_{44} > 0$, $C_{11} - C_{12} > 0$, and $C_{11} + 2C_{12} + 4C_{44} > 0$. Focusing on the negative stiffness $E < 0$ scenario, one can derive the following requirements from these inequalities,

$$E < 0, \ \nu < -1, \ G > 0, \ \text{and} \ \frac{E}{1-2\nu} + 4G > 0. \tag{S14}$$

Note that the first three requirements are the same for an isotropic material, and the additional requirement, $\frac{E}{1-2\nu} + 2G > 0$, arises because of the direction dependency. Eq. (S14) can be used as a necessary condition to find cubic metamaterials with negative stiffness that have a great chance of satisfying strong ellipticity. Once a promising candidate is found, the procedure described in Fig. S6 can be used to give a strict validation of the strong ellipticity condition.

### Elastic parameters in arbitrary loading directions

Based on the cubic symmetry, the elastic compliant tensor $S_{ijkm}$ of a cubic material can be written in the following form [15],

$$S_{ijkm}^{'} = S_{12}\delta_{ij}\delta_{km} + \frac{1}{4}S_{44}(\delta_{ik}\delta_{jm} + \delta_{im}\delta_{jk}) + (S_{11} - S_{12} - \frac{1}{4}S_{44})a_{it}a_{jt}a_{kt}a_{mt}. \tag{S15}$$

where $S_{ijkm}$ denotes the compliant tensor in the original lattice structure and $'$ denotes the compliant tensor in the new coordinate system defined by $[hkl]$ and $\theta$. The notation $a_{it}a_{jt}a_{kt}a_{mt}$ is adopted from [15], which equals to $\sum_{t=1,2,3} a_{it}a_{jt}a_{kt}a_{mt}$. The transformation matrix is

$$a_{st} = \begin{bmatrix} \frac{h}{\sqrt{h^2+k^2+l^2}} & \frac{k}{\sqrt{h^2+k^2+l^2}} & \frac{l}{\sqrt{h^2+k^2+l^2}} \\ -\frac{\sqrt{k^2+l^2}\text{Cos}[\theta]}{\sqrt{h^2+k^2+l^2}} & \frac{hk\text{Cos}[\theta]}{\sqrt{k^2+l^2}\sqrt{h^2+k^2+l^2}} - \frac{l\text{Sin}[\theta]}{\sqrt{l^2+k^2}} & \frac{hl\text{Cos}[\theta]}{\sqrt{k^2+l^2}\sqrt{h^2+k^2+l^2}} + \frac{k\text{Sin}[\theta]}{\sqrt{l^2+k^2}} \\ \frac{\sqrt{k^2+l^2}\text{Sin}[\theta]}{\sqrt{h^2+k^2+l^2}} & -\frac{l\text{Cos}[\theta]}{\sqrt{l^2+k^2}} - \frac{hk\text{Sin}[\theta]}{\sqrt{k^2+l^2}\sqrt{h^2+k^2+l^2}} & \frac{k\text{Cos}[\theta]}{\sqrt{h^2+k^2}} - \frac{hl\text{Sin}[\theta]}{\sqrt{k^2+l^2}\sqrt{h^2+k^2+l^2}} \end{bmatrix}. \quad \text{(S16)}$$

In order to derive the compliant matrix, a transformation between matrix notation and tensor notation should be performed first. Using $S_{11}^{'} = S_{1111}^{'}$, $S_{12}^{'} = S_{1122}^{'}$, $S_{44}^{'} = 4S_{2323}^{'}$, the following equations are derived from Eq. (S15)

$$S_{11}^{'} = S_{12} + (S_{11} - S_{12} - \frac{1}{2}S_{44})(1 - a_{1t}a_{1t}a_{1t}a_{1t}) \quad \text{(S17a)}$$

$$S_{12}^{'} = S_{12} + (S_{11} - S_{12} - \frac{1}{2}S_{44})a_{1t}a_{2t}a_{1t}a_{2t} \quad \text{(S17b)}$$

$$S_{44}^{'} = S_{44} + 4(S_{11} - S_{12} - \frac{1}{2}S_{44})a_{2t}a_{3t}a_{2t}a_{3t} \quad \text{(S17c)}$$

All these equations can be written in a general form as Eq. (3) in the main text,

$$S_{ij}^{'} = S_{ij} + A \cdot F_{ij}([hkl], \theta). \quad \text{(S18)}$$

where $A = (S_{11} - S_{12} - \frac{1}{2}S_{44})$ is the anisotropic parameter. $F_{ij}([hkl], \theta)$ is a function of direction vector $[hkl]$ and rotation angle $\theta$. The expression of $F_{ij}$ is calculated by combing Eq. (S16) and (S17) which are listed in the following

$F_{11} = -2(h^2k^2 + k^2l^2 + l^2h^2)$

$F_{12} = \frac{2}{k^2+l^2}(h^2(k^4 + k^2l^2 + l^4)\text{Cos}[\theta]^2 + hkl(-k^2 + l^2)\text{Cos}[\theta]\text{Sin}[\theta] + k^2l^2\text{Sin}[\theta]^2)$

$F_{13} = \frac{2}{k^2+l^2}(k^2l^2\text{Cos}[\theta]^2 + hkl(k^2 - l^2)\text{Cos}[\theta]\text{Sin}[\theta] + h^2(k^4 + k^2l^2 + l^4)\text{Sin}[\theta]^2)$

$F_{44} = \frac{4}{(k^2+l^2)^2}[(k^2+l^2)^4\text{Cos}[\theta]^2\text{Sin}[\theta]^2 + (l\text{Cos}[\theta] + hk\text{Sin}[\theta])^2(hk\text{Cos}[\theta] - l\text{Sin}[\theta])^2$

$\qquad + (hl\text{Cos}[\theta] + k\text{Sin}[\theta])^2(k\text{Cos}[\theta] - hl\text{Sin}[\theta])^2]$

The stiffness of a cubic material in an arbitrary direction $[hkl]$ can then be calculated from

$$\frac{1}{E^{'}} = \frac{1}{E_{[hkl]}} = S_{11}^{'} = \frac{1}{E_{[100]}} - 2A(h^2k^2 + k^2l^2 + l^2h^2). \quad \text{(S19)}$$

Eq. (S19) recovers the elastic representative surface of a cubic material as in [16]. Note that $E_{[hkl]}$ only depends on the direction of compression $[hkl]$.

Using a similar approach, Poisson's ratio can be derived as

$$\nu' = \nu_{[hkl],\theta} = -\frac{S_{12}'}{S_{11}'} = -\frac{S_{12} + A \cdot F_{12}([hkl],\theta)}{S_{11} + A \cdot F_{11}([hkl],\theta)}. \tag{S20}$$

Since Poisson's ratio depends on both $[hkl]$ and $\theta$, a full picture is tricky to be visualized. Here, the stereographic projection in crystallography is adopted to visualize how Poisson's ratio evolves under different loading directions [Fig. S7(c) and Fig. 3(c)]. In addition, the average value of $\nu'$ on the $[hkl]$ plane can be calculated by integrating over angle $\theta$,

$$\overline{S}_{ij}' = \frac{1}{2\pi}\int_0^{2\pi}\left(S_{ij} + AF_{ij}([hkl],\theta)\right)d\theta = S_{ij} + A\cdot\frac{1}{2\pi}\int_0^{2\pi}F_{ij}([hkl],\theta)d\theta = S_{ij} + A\cdot\overline{F}_{ij}([hkl]). \tag{S21}$$

Average of $F_{ij}$ is are given in the following,

$$\overline{F}_{11} = F_{11}, \tag{S22}$$

$$\overline{F}_{12} = \frac{1}{k^2+l^2}(k^2l^2 + h^2(k^4 + k^2l^2 + l^4)), \tag{S23}$$

$$\overline{F}_{13} = \frac{1}{k^2+l^2}(k^2l^2 + h^2(k^4 + k^2l^2 + l^4)), \tag{S24}$$

$$\overline{F}_{44} = \frac{1}{2(k^2+l^2)^2}((k^2+l^2)^4 + (h^2k^2+l^2)^2 + (k^2+h^2l^2)^2) \tag{S25}$$

An elastic representative surface of $\overline{\nu'}$ is also plotted in Fig. 3(c) to visualize the direction dependency.

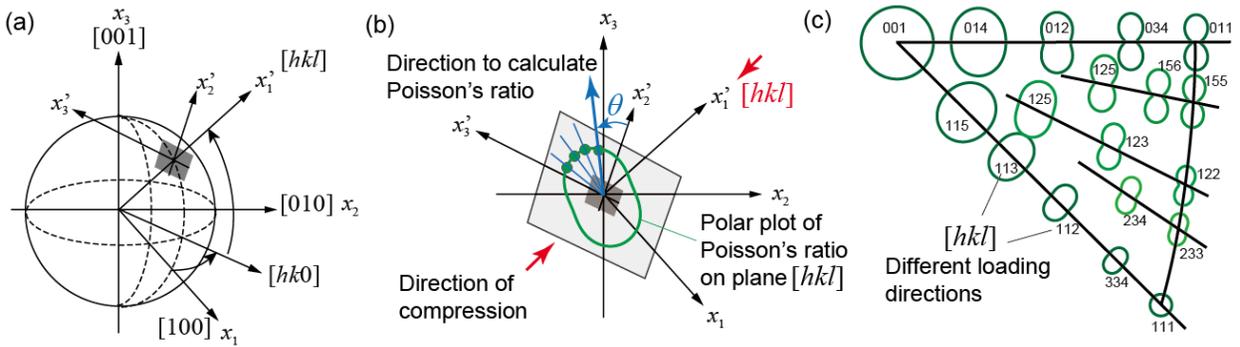

FIG. S7 Schematics illustrating the Poisson's ratio of a 3D material on an arbitrary loading direction and direction, which are defined by $[hkl]$ and $\theta$. On each loading direction, a polar plot is plotted with varying $\theta$. The stereographic projection in crystallography is then adopted to represent the dependency of Poisson's ratio on both $[hkl]$ and $\theta$.

**Pattern transformation**

Cross section views on the (011) plane are provided in Fig. S8 to highlight the anti-symmetric to symmetric pattern transformation. The deformation contours of 1 conventional unit cell, 2×2×2, 3×3×3, 1×1×4, and 2×2×4 conventional unit cells under uniaxial compression are provided in Fig. S9. Numerical results validate that the conventional unit cell captures the critical pattern. To demonstrate the effect of imperfection, buckling modes are introduced as imperfection, and the effect of imperfections are shown in Fig. S10. The effect of hyperelasticity is presented in Fig. S11 using an Arruda-Boyce hyperelastic material model. It also should be pointed out that pressure inside the spheres will also have an effect on the mechanical behavior of the material, especially when it's inflated to high pressure and the shells are made of compliant materials. However, at atmospheric pressure, this effect is insignificant, since the deformation energy of the shells is much greater than the compression energy of gases. On the other hand, pressure can be artificially introduced to the spheres to achieve active control of the metamaterial, which is an interesting topic that can be further explored.

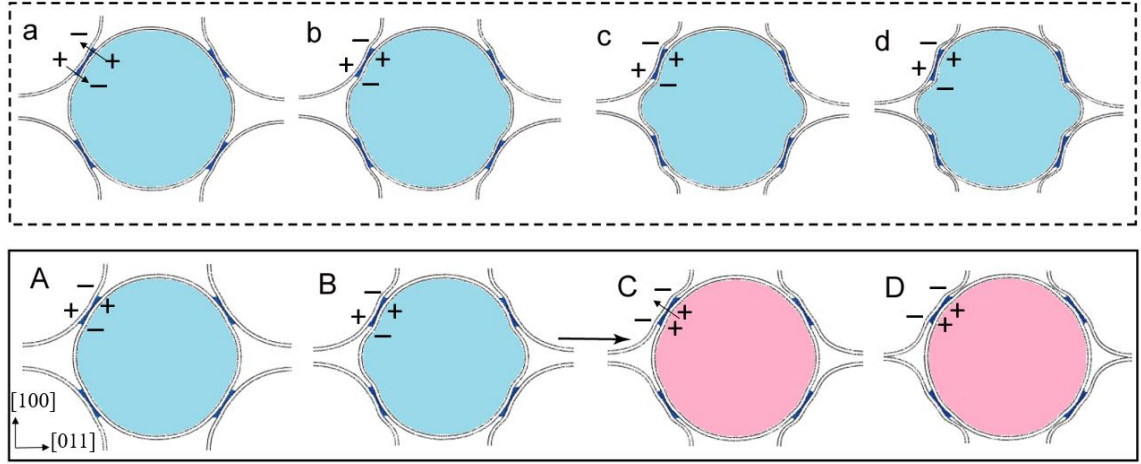

FIG. S8 Supplementary figure to Fig. 4 in the main text. a-d and A-D plots the deformation of the metamaterial in cross section on the plane (011). Note compressive strain is applied in the [100] direction. The "+" and "−" sign is used to highlight the rise and dents of the spherical shells. The blue color highlights the anti-symmetric deformation of the two spheres in contact. The pint color highlights the non-symmetric deformation of the two spheres in contact. A pattern transformation from anti-symmetric deformation to non-symmetric deformation is observed in the bottom case.

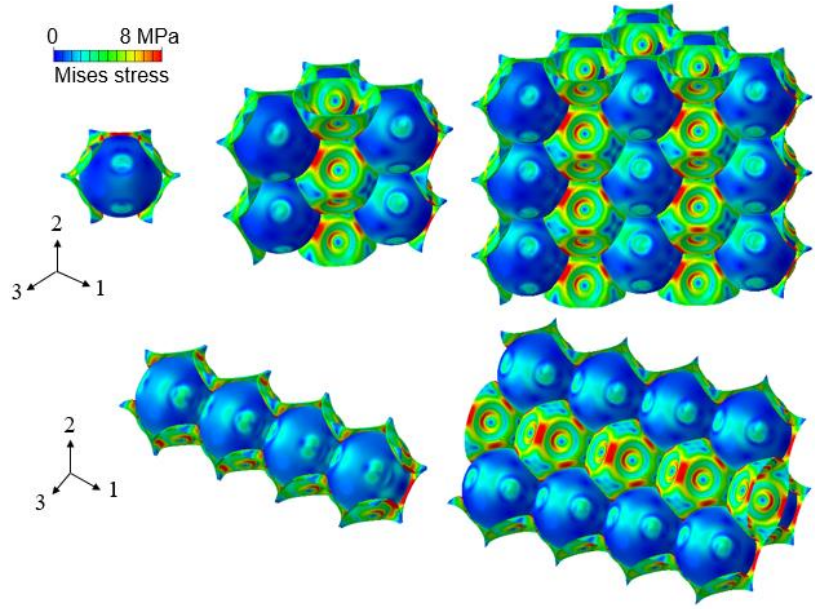

FIG. S9 Validating the buckling pattern in multiple unit cells. The results of 1 conventional BCC unit cell, 2×2×2, 3×3×3, 1×1×4, and 2×2×4 conventional unit cells are presented at $\varepsilon_{11} = -0.15$. Part of the spheres is removed to see the deformation pattern. The material has $h/R = 0.02$, $E_b/E_s = 0.04$, $r_b/R = 0.2$ and $t/R = 0.04$.

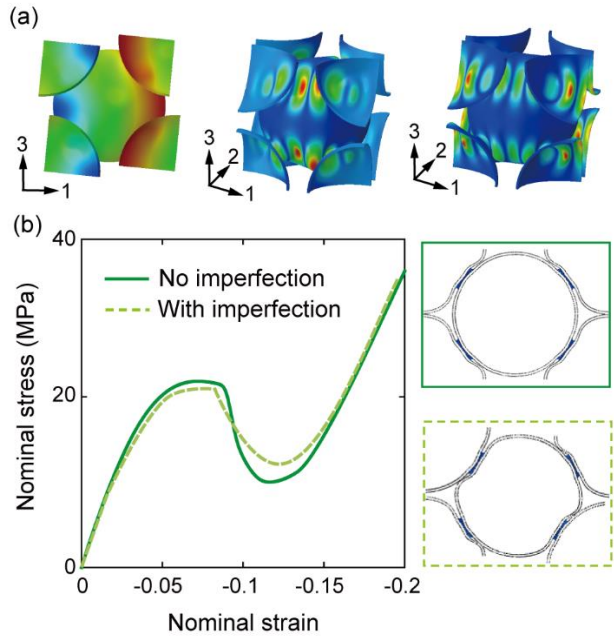

FIG. S10 Effect of imperfection. (a) shows the three eigen-modes introduced to the material as imperfections. (b) shows the effect of imperfection on the stress-strain curve and possible buckling patterns.

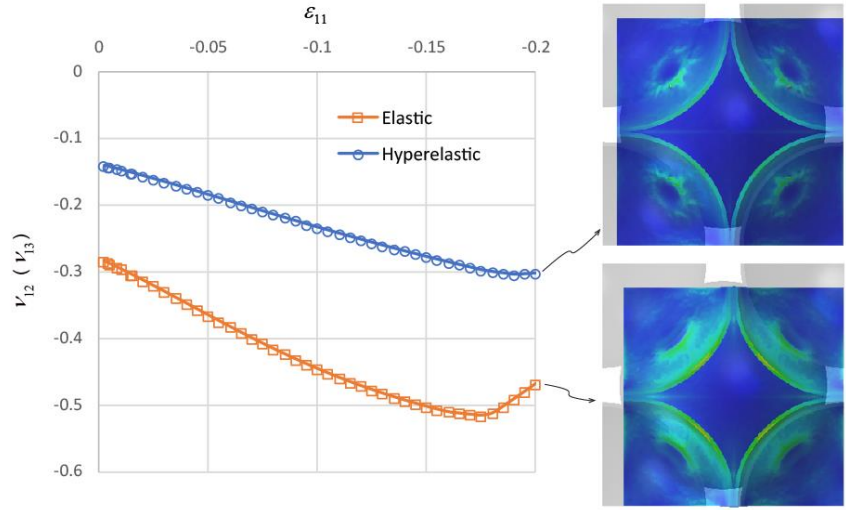

FIG. S11 Comparison between a linear elastic material and a hyperelastic material. The Hyperelastic material is model with the Arruda-Boyce model, with initial shear modulus 0.18 MPa, locking stretch of 2.05, and bulk modulus of 1 GPa. Same material is used for both binders and shells. The geometry parameters used are $r_b/R = 0.2$, $h/R = 0.02$, and $t/R = 0.04$.

---

[1] Z. Wang, T. Zhang, B. K. Park, W. I. Lee, and D. J. Hwang, J. Mater. Sci., 1 (2016).

[2] W. Sanders and L. Gibson, Mater. Sci. Eng. A **347**, 70 (2003).

[3] Z. Jia, Z. Wang, D. Hwang, and L. Wang, ACS Applied Energy Materials (2018).

[4] M. Danielsson, D. Parks, and M. Boyce, J. Mech. Phys. Solids. **50**, 351 (2002).

[5] L. Wang, M. C. Boyce, C. Y. Wen, and E. L. Thomas, Adv. Funct. Mater. **19**, 1343 (2009).

[6] E. Riks, Int. J. Solids Struct. **15**, 529 (1979).

[7] L. Rothenburg, A. A. Berlin, and R. J. Bathurst, Nature **354**, 470 (1991).

[8] I. Shufrin, E. Pasternak, and A. V. Dyskin, Int. J. Solids Struct. **54**, 192 (2015).

[9] R. S. Lakes, T. Lee, A. Bersie, and Y. Wang, Nature **410**, 565 (2001).

[10] G. Geymonat, S. Müller, and N. Triantafyllidis, Archive for rational mechanics and analysis **122**, 231 (1993).

[11] D. Han, H.-H. Dai, and L. Qi, Journal of Elasticity **97**, 1 (2009).

[12] J. L. Rose, *Ultrasonic waves in solid media* (Cambridge university press, 2004).

[13] S. Xinchun and R. S. Lakes, Physica status solidi (b) **244**, 1008 (2007).

[14] G. F. Miller and M. Musgrave, Proceedings of the Royal Society of London. Series A. Mathematical and Physical Sciences **236**, 352 (1956).

[15] T. Thomas, Proc. Natl. Acad. Sci. **55**, 235 (1966).

[16] J. F. Nye, *Physical properties of crystals: their representation by tensors and matrices* (Oxford university press, 1985).